\documentclass[runningheads]{llncs}
\usepackage{amssymb,amsmath}
\usepackage{pstricks}
\usepackage{pst-node}
\usepackage{epsfig}
\usepackage{latexsym}
\begin{document}
\pagestyle{headings}

\newcommand{\Implies}{\rightarrow}
\newcommand{\Equiv}{\leftrightarrow}
\let\goodoldAnd=\And
\renewcommand{\And}{\wedge}
\newcommand{\Or}{\vee}
\newcommand{\Not}{\neg}
\newcommand{\Falsum}{\bot}
\newcommand{\Verum}{\top}
\newcommand{\If}{\leftarrow}

\newcommand{\Nat}{\ensuremath{\mathbb{N}}}

\title{%
       An Almost Classical Logic for
       Logic Programming and
       Nonmonotonic Reasoning 
\thanks{Originally published in proc. PCL 2002, a FLoC workshop;
eds. Hendrik Decker, Dina Goldin, J{\o}rgen Villadsen, Toshiharu Waragai
({\tt http://floc02.diku.dk/PCL/}).}
}

\titlerunning{An Almost Classical Logic}

\author{%
        Fran\c{c}ois Bry
       }

\institute{
           Institute for Computer Science, University of Munich, Germany\\
           \email{http://www.pms.informatik.uni-muenchen.de/}
          }

\maketitle

\begin{abstract}
  
  The model theory of a first-order logic called N$^4$ is introduced.
  N$^4$ does not eliminate double negations, as classical logic does,
  but instead reduces fourfold negations. N$^4$ is very close to
  classical logic: N$^4$ has two truth values; implications are, in
  N$^4$ like in classical logic, material; and negation distributes
  over compound formulas in N$^4$ as it does in classical logic.
  Results suggest that the semantics of normal logic programs is
  conveniently formalized in N$^4$: Classical logic Herbrand
  interpretations generalize straightforwardly to N$^4$; the classical
  minimal Herbrand model of a positive logic program coincides with
  its unique minimal N$^4$ Herbrand model; the stable models of a
  normal logic program and its so-called complete minimal N$^4$
  Herbrand models coincide.

\end{abstract}

\section{Introduction}

This paper first introduces the (classical style) model theory of a
first-order logic called N$^4$.  The salient characteristic of N$^4$
is that it does not eliminate double negations as classical logic does
(in N$^4$, $\Not \Not F$ is not logically equivalent to $F$ and $\Not
\Not \Not F$ is not logically equivalent to $\Not F$), but instead it
reduces fourfold negations (in N$^4$, $\Not \Not \Not \Not F$ is
logically equivalent to $\Not \Not F$). The name N$^4$ stresses that
fourfold negations are reduced.

Despite its nonstandard treatment of negation, N$^4$ is very close to
classical logic. Like classical logic, N$^4$ has two truth values, its
implication is material (in N$^4$, $A \Implies B$ is logically
equivalent to $\Not A \Or B$), and the truth value of a formula is
defined recursively in terms of the truth values of its subformulas.
Most logical consequences of classical logic hold also in N$^4$. In
particular, negation distributes over compound formulas in N$^4$ as it
does in classical logic. Also, in N$^4$ $F$ logically implies $\Not
\Not F$ (but the converse does not hold) and three laws of excluded
middle hold ($F \Or \Not F$, $\Not F \Or \Not \Not F$, and $\Not \Not
F \Or \Not \Not \Not F$ are always true but $F \Or \Not \Not \Not F$
might be false in some so-called incomplete N$^4$ interpretations).
Furthermore, a classical logic model of a set $\cal S$ of formulas is
also a N$^4$ model of $\cal S$.

This paper investigates formalizing the semantics of normal logic
programs using N$^4$. A few results suggest that N$^4$ is convenient
for this purpose. Classical logic Herbrand interpretations generalize
straightforwardly to N$^4$ as interpretations characterized by the
ground atoms and the doubly negated ground atoms (instead of only the
ground atoms) they satisfy. The classical minimal Herbrand model of a
positive logic program coincides with its (unique) minimal N$^4$
Herbrand model. Every normal logic program has (in general many)
minimal N$^4$ Herbrand models. The stable models of a normal logic
program \cite{gelfond88} coincide with its so-called complete minimal
N$^4$ Herbrand models.

This paper is structured as follows. The next section, Section 2,
recalls a few syntax notions and introduces terminology and notations.
Section 3 defines the model theory of N$^4$ and gives a few results on
logical consequence in N$^4$. Section 4 is devoted to N$^4$ Herbrand
interpretations. In Section 5, the minimal N$^4$ Herbrand models of
normal logic programs are defined and investigated. Section 6
discusses the intuitive meaning of N$^4$.  Section 7 addresses
perspectives and related work. The proofs are given in Appendix.

\section{Syntax, Terminology, and Notations}

N$^4$ syntax is that of classical first-order logic. 
If $\cal L$ is a first-order language, 
its (non-empty) set of constants will be noted {\it Const}$_{\cal L}$, 
the set of its $n$-ary $(n \geq 1)$ function symbols will be noted 
   {\it Fun}$^n_{\cal L}$,
and the set of its $n$-ary $(n \geq 0)$ predicate symbols will be noted 
   {\it Rel}$^n_{\cal L}$. 
A first-order language is assumed to include 
the {\em falsum} $\Falsum$, the unary connective $\Not$, 
the binary connectives $\And$ $\Or$, 
and the quantifiers $\forall$ $\exists$.

In the following, a fixed first-order language $\cal L$ is assumed. 
The terms, ground terms, Herbrand universe, atoms or atomic formulas,
formulas, closed formulas, etc.\ of $\cal L$ are defined
as usual. Note that the falsum $\Falsum$ is not an atom. $n$-fold $(n
\geq 0)$ negations will be noted $\Not^n$. A formula $F$ is in 
{\em prefix negation form} if $F = \Not^n G$ $(n \geq 0)$ and no
negations occur in $G$. 
Two additional connectives, $\Implies$ and $\Equiv$, and the {\em verum}
$\Verum$ are defined as follows as shorthand notations: 
$(F \Implies G) := (\Not F \Or G)$, 
$(F \Equiv G)   := ((\Not F \Or G) \And (\Not G \Or F))$, 
and $\Verum         := \Not \Falsum$.

The following unusual notion of literal will be used. 
\begin{definition}[N$^4$ literal]
A {\em {\rm N}$^4$ literal} is an atom, a negated atom, a doubly negated
atom, or a threefold negated atom. A {\em positive {\rm N}$^4$
  literal} is an atom or a doubly negated 
atom. A {\em negative {\rm N}$^4$ literal} is a negated atom or a
threefold negated atom. 
\end{definition}

A {\em positive program clause} ({\em general program clause}, resp.)
in $\cal L$ is an expression of the form $A \If B_1, \ldots, B_n$ ($n
\geq 0$), where $A$ is an atom of $\cal L$ and $B_1, \ldots,$ and
$B_n$ are atoms (atoms  or negated atoms, resp.) of $\cal L$. A 
{\em positive} ({\em normal or general,} resp.) logic program in
${\cal L}$ is a finite set of positive (general, resp.) 
program clauses in $\cal L$. 

\section{N$^4$ Model Theory}

N$^4$ interpretations resemble that of classical logic. A 
significant difference is that they assign
relations not only to predicate symbols, as classical logic interpretations
do, but also to doubly negated predicate symbols. 

\begin{definition}[N$^4$ Interpretation]\label{def:n3-interpretation}
A {\em N$^4$ interpretation} $\cal I$ of $\cal L$ is a pair 
$(D_{\cal I}, {\it val}_{\cal I})$ such that 
\begin{enumerate}
   \item $D_{\cal I}$ is a non-empty set, called the {\em domain} or
         {\em universe} of $\cal I$.
   \item ${\it val}_{\cal I}$ is an assignment defined as follows: 
   \begin{enumerate}
      \item[2.1] ${\it val}_{\cal I}(c) \in D_{\cal I}$, for 
                 $c \in${\it Const}$_{\cal L}$.
      \item[2.2] ${\it val}_{\cal I}(f)$ is a function from $D_{\cal I}^{~n}$
                 into $D_{\cal I}$, for $f \in {\it Fun}^n_{\cal L}$ $(n \geq 1)$. 
      \item[2.3] ${\it val}_{\cal I}(p) \in \{${\bf true}, {\bf false}$\}$
                 and 
                 ${\it val}_{\cal I}(\Not^2 p) \in 
                  \{${\bf true}, {\bf false}$\}$ 
                 such that
                 if ${\it val}_{\cal I}(p) = {\bf true}$, 
                 then 
                 ${\it val}_{\cal I}(\Not^2 p) = {\bf true}$,
                 for $p \in {\it Rel}^0_{\cal L}$. 
      \item[2.4] ${\it val}_{\cal I}(p) \subseteq 
                  {\it val}_{\cal I}(\Not^2 p) \subseteq 
                  D_{\cal I}^{~n}$,
                 for $p \in {\it Rel}^n_{\cal L}$ $(n \geq 1)$.
   \end{enumerate}
\end{enumerate}
\end{definition}
Thus, in a N$^4$ interpretation only one of the three truth
assignments of Figure \ref{fig:first-truth-values-table} are possible for
a propositional variable $p$ and the positive N$^4$ literal $\Not^2
p$. Note that if $p$ is true, then $\Not^2 p$ is also true. 

A classical logic interpretation trivially induces a N$^4$
interpretation: It suffices to assign the same truth value or
relation to each doubly negated predicate symbol $\Not^2 p$ as to
$p$. However, not every N$^4$ interpretation corresponds to a
classical logic interpretation. For example, the second line of 
Figure \ref{fig:first-truth-values-table} is not possible in 
classical logic.

\begin{figure}
\begin{center}
\begin{tabular}{|c|c|}
\hline
$p$         & $\Not^2 p$ \\
\hline\hline
{\bf true}  & {\bf true} \\
\hline
{\bf false} & {\bf true} \\
\hline
{\bf false} & {\bf false} \\
\hline
\end{tabular}
\end{center}
\vspace{- 1em}
\caption{Possible valuations of $p$ and $\Not^2 p$ in N$^4$
  interpretations}\label{fig:first-truth-values-table} 
\end{figure}

In N$^4$, variable and term assignments are defined like in
classical logic. Both definitions are recalled, so as to introduce
the notations used later.
 
\begin{definition}[Variable Assignment]
Let $\cal I$ be a {\rm N}$^4$ interpretation of $\cal L$ with domain
$D_{\cal I}$. A {\em variable assignment}  $\cal V$ 
with respect to $\cal I$ assigns an element of $D_{\cal I}$
to each variable of $\cal L$. 
If $\cal V$ is variable assignment with respect to 
$\cal I$, $x$ is a variable, and $d \in D_{\cal I}$, then 
${\cal V}\left[ d / x \right]$ denotes the following variable
assignment with respect to $\cal I$:  

\vspace{-1 em}

\[ {\cal V}\left[ d/x \right](y) := 
     \left\{ \begin{array}{l@{\quad}l}
                          d           & {\rm if~} y = x\\
                          {\cal V}(y) & {\rm otherwise}
              \end{array} 
     \right. 
\]

\end{definition}

\begin{definition}[Term Assignment]
Let $\cal I = (D_{\cal I}, {\it val}_{\cal I})$ be a {\rm N}$^4$
interpretation of $\cal L$
and $\cal V$ a variable assignment with respect to $\cal I$. 
The {\em term assignment} ${\it val}_{\cal I, V}$ 
with respect to $\cal I$ and $\cal V$ is defined as follows: 
\begin{enumerate}
   \item ${\it val}_{\cal I, V}(x) = {\cal V}(x)$, for $x$ variable.
   \item ${\it val}_{\cal I, V}(c) = {\it val}_{\cal I}(c)$, for 
          $c \in$ {\it Const}$_{\cal L}$.
   \item ${\it val}_{\cal I, V}(f(t_1, \ldots, t_n)) = 
         {\it val}_{\cal I}(f)({\it val}_{\cal I, V}(t_1), 
            \ldots, {\it val}_{\cal I, V}(t_n))$, 
         for $f \in {\it Fun}^n_{\cal L}$ $(n \geq 1)$ 
         and $t_1, \ldots t_n$ terms. 
\end{enumerate}
\end{definition}

\begin{figure}
\begin{center}
\begin{tabular}{|c|c|c|c|}
\hline
$p$         & $\Not p$    & $\Not^2 p$ & $\Not^3 p$ \\
\hline\hline
{\bf true}  & {\bf false} & {\bf true}    & {\bf false}\\
\hline
{\bf false} & {\bf true} & {\bf true}     & {\bf false}\\
\hline
{\bf false} & {\bf true} & {\bf false}    & {\bf true}\\
\hline
\end{tabular}
\end{center}
\vspace{- 1em}
\caption{Possible valuations of $p$, $\Not p$, 
         $\Not^2 p$, and $\Not^3 p$ in N$^4$ 
         interpretations}\label{fig:second-truth-values-table} 
\end{figure}

The truth value of a negated formula is
defined in such a way, that Figure \ref{fig:first-truth-values-table} can
be completed as shown by Figure \ref{fig:second-truth-values-table}, i.e.\
$\Not p$ negates $p$ but not $\Not^2 p$ and $\Not^3 p$
negates $\Not^2 p$ but not $p$.  

\begin{definition}[Formula Valuation]\label{def:formula-valuation}
Let $\cal I = (D_{\cal I}, {\it val}_{\cal I})$ be a 
{\rm N}$^4$ interpretation of $\cal L$ and
$\cal V$ a variable assignment with respect to $\cal I$. The valuation
function $val_{\cal I, V}$ with respect to $\cal I$ and $\cal V$ is 
defined as follows: 
\begin{enumerate}
   \item[1.1] $val_{\cal I, V}(p) = {\bf true}$ iff 
              ${\it val}_{\cal I}(p) = {\bf true}$, 
              for $p \in {\it Rel}^0_{\cal L}$ and \\
              $val_{\cal I, V}(p(t_1, \ldots, t_n)) = {\bf true}$ iff 
              $({\it val}_{\cal I, V}(t_1), \ldots, {\it val}_{\cal I, V}(t_n)) \in 
                 {\it val}_{\cal I}(p)$, 
              for $p \in {\it Rel}^n_{\cal L}$ $(n \geq 1)$ and 
              $t_1, \ldots, t_n$ terms.
   \item[1.2] $val_{\cal I, V}((F_1 \And F_2)) =  {\bf true}$ iff 
              $val_{\cal I, V}(F_1) = {\bf true}$ and 
              $val_{\cal I, V}(F_2) = {\bf true}$.
   \item[1.3] $val_{\cal I, V}((F_1 \Or F_2)) = {\bf true}$ iff 
              $val_{\cal I, V}(F_1) = {\bf true}$ or
              $val_{\cal I, V}(F_2) = {\bf true}$.
   \item[1.4] $val_{\cal I, V}(\forall x F) = {\bf true}$ iff 
              for all $d \in D_{\cal I}$
                 $val_{{\cal I},{\cal V}[d/x]}(F) = {\bf true}$.
   \item[1.5] $val_{\cal I, V}(\exists x F) = {\bf true}$ iff 
              for some $d \in D_{\cal I}$
                 $val_{{\cal I},{\cal V}[d/x]}(F) = {\bf true}$.

   \item[2.1] $val_{\cal I, V}(\Not \Falsum) =$ {\bf true}, \\
              $val_{\cal I, V}(\Not p) =$ {\bf true} iff 
              $val_{\cal I, V}(p)   \neq$ {\bf true}, 
              for $p \in {\it Rel}^n_{\cal L}$ $(n \geq 0)$, and
              $val_{\cal I, V}(\Not p(t_1, \ldots, t_n)) = {\bf true}$ iff 
              $({\it val}_{\cal I, V}(t_1), \ldots, {\it val}_{\cal I, V}(t_n)) \not\in 
                 {\it val}_{\cal I}(p)$, 
              for $p \in {\it Rel}^n_{\cal L}$ $(n \geq 1)$.              
   \item[2.2] $val_{\cal I, V}(\Not (F_1 \And F_2)) = 
               val_{\cal I, V}((\Not F_1 \Or  \Not F_2))$.
   \item[2.3] $val_{\cal I, V}(\Not(F_1 \Or   F_2)) = 
               val_{\cal I, V}((\Not F_1 \And \Not F_2))$.
   \item[2.4] $val_{\cal I, V}(\Not \forall x F) = 
               val_{\cal I, V}(\exists x \Not F)$.
   \item[2.5] $val_{\cal I, V}(\Not \exists x F) = 
               val_{\cal I, V}(\forall x \Not F)$.

   \item[3.1] $val_{\cal I, V}(\Not^2 p) = {\bf true}$ iff 
              ${\it val}_{\cal I}(\Not^2 p) = {\bf true}$, 
              for $p \in {\it Rel}^0_{\cal L}$ and\\
              $val_{\cal I, V}(\Not^2 p(t_1, \ldots, t_n)) = 
                 {\bf true}$ iff 
              $({\it val}_{\cal I, V}(t_1), \ldots, {\it val}_{\cal I, V}(t_n)) \in$ 
                 ${\it val}_{\cal I}(\Not^2 p)$, 
              for $p \in {\it Rel}^n_{\cal L}$ $(n \geq 1)$ and 
              $t_1, \ldots, t_n$ terms.
   \item[3.2] $val_{\cal I, V}(\Not^2 (F_1 \And F_2)) = 
               val_{\cal I, V}((\Not^2 F_1 \And \Not^2 F_2))$.
   \item[3.3] $val_{\cal I, V}(\Not^2(F_1 \Or F_2)) = 
               val_{\cal I, V}((\Not^2 F_1 \Or \Not^2 F_2))$.
   \item[3.4] $val_{\cal I, V}(\Not^2 \forall x F) = 
               val_{\cal I, V}(\forall x \Not^2 F)$.
   \item[3.5] $val_{\cal I, V}(\Not^2 \exists x F) = 
               val_{\cal I, V}(\exists x \Not^2 F)$.

   \item[4~~] $val_{\cal I, V}(\Not^3 F) =$ {\bf true} iff 
              $val_{\cal I, V}(\Not^2 F) \neq$ {\bf true}.

   \item[5~~] $val_{\cal I, V}(F) = {\bf false}$ iff 
              $val_{\cal I, V}(F) \neq {\bf true}$.
\end{enumerate}
\end{definition}
On can prove as follows that $val_{\cal I, V}$ is a total
function over the formulas of $\cal L$. For
each formula $F$ exactly one of the clauses 1.1 to 4 of Definition
\ref{def:formula-valuation} apply. (Which clause applies to a formula
depends on its structure.) Therefore, 
$val_{\cal I, V}(.) = {\bf true}$ defines a partial function. Because 
of clause 5, $val_{\cal I, V}$ is total. Since $val_{\cal I, V}$ is a
total function, Definition \ref{def:formula-valuation} correctly specifies
the valuation of formulas in an N$^4$ interpretation. 

Definition \ref{def:formula-valuation} differs from its classical logic
counterpart as follows. To obtain the definition of classical
logic, drop clauses 2.1 through 3.5 and replace clause 4 by:  
\begin{enumerate}
   \item[{\it 4'}~] $val_{\cal I, V}(\Not F) = {\bf true}$ {\it iff}
                    $val_{\cal I, V}(F) \neq {\bf true}$.
\end{enumerate}
Note that, although clauses 2.1 through 3.5 are
not needed in the classical logic counterpart of Definition
\ref{def:formula-valuation}, they hold in classical logic. Note also that
the elimination of double negation follows from clause 4'.  

Properties of N$^4$ are given in the rest of this
section. From now on, $\cal I = (D_{\cal I},  
{\it val}_{\cal I})$ denotes a N$^4$ interpretation of 
$\cal L$, $\cal V$ a variable assignment with respect to $\cal I$, and $F$,
$F_1$, $F_2$, $F_3$, $G$, $G_1$, and $G_2$ formulas of $\cal L$. 

\begin{proposition}\label{prop:replacement}~
\begin{enumerate}
   \item $val_{\cal I, V}(\Falsum)      = {\bf false}$ and 
         $val_{\cal I, V}(\Verum) = {\bf true}$
   \item $val_{\cal I, V}((F_1 \And F_2)) = 
          val_{\cal I, V}((F_2 \And F_1))$.
   \item $val_{\cal I, V}((F_1 \Or F_2)) = 
          val_{\cal I, V}((F_2 \Or F_1))$.
   \item $val_{\cal I, V}((F_1 \And (F_2 \Or F_3))) = 
          val_{\cal I, V}(((F_1 \And F_2) \Or (F_1 \And F_3)))$.
   \item $val_{\cal I, V}((F_1 \Or (F_2 \And F_3))) = 
          val_{\cal I, V}(((F_1 \Or F_2) \And (F_1 \Or F_3)))$.
   \item $val_{\cal I, V}((F_1 \And (F_2 \And F_3))) = 
          val_{\cal I, V}(((F_1 \And F_2) \And F_3))$.
   \item $val_{\cal I, V}((F_1 \Or (F_2 \Or F_3))) = 
          val_{\cal I, V}(((F_1 \Or F_2) \Or F_3))$.
   \item If $val_{\cal I, V}(G_1) = val_{\cal I, V}(G_2)$, then 
         \begin{enumerate}
            \item[] $val_{\cal I, V}((F \Or G_1)) = 
                     val_{\cal V, I}((F \Or G_2))$
            \item[] $val_{\cal I, V}((F \And G_1)) = 
                     val_{\cal V, I}((F \And G_2))$
            \item[] $val_{\cal I, V}(\Not G_1) = val_{\cal V, I}(\Not G_2)$
         \end{enumerate}
   \item $val_{\cal I, V}(\Not \forall x F) = val_{\cal I, V}(\exists x \Not F)$
   \item $val_{\cal I, V}(\Not \exists x F) = val_{\cal I, V}(\forall x \Not F)$

\end{enumerate}
\end{proposition}

\begin{proposition}\label{prop:p-not-p}
If $F$ is in prefix negation form and for all $k \in \Nat$
$F \neq \Not^{2k+1} G$, then 
$val_{\cal I, V}(F) \neq val_{\cal I, V}(\Not F)$.
\end{proposition}

As Figures \ref{fig:second-truth-values-table} and 
\ref{fig:excluded-middle-falsifying-interpretation} show, Proposition
\ref{prop:p-not-p} does not generalize to all formulas. 

\begin{proposition}\label{prop:excluded-middle}~
\begin{enumerate}
  \item {\em Fourfold negation reduction:}
         $val_{\cal I, V}(\Not^4 F) = val_{\cal I, V}(\Not^2 F)$

   \item {\em Laws of excluded middle:}\\
         $val_{\cal I, V}((F \Or \Not F)) = 
          val_{\cal I, V}((\Not F \Or \Not^2 F)) =
          val_{\cal I, V}((\Not^2 F \Or \Not^3 F)) =
         {\bf true}$

   \item {\em Laws of excluded contradiction:}\\
         $val_{\cal I, V}((\Not F \And \Not^2 F)) = 
          val_{\cal I, V}((\Not^2 F \And \Not^3 F)) = 
          val_{\cal I, V}((F \And \Not^3 F)) = 
         {\bf false}$
\end{enumerate}
\end{proposition}

\begin{figure}
\begin{center}
\begin{tabular}{|c|c|c|c|}
\hline
$p$         & $\Not p$   & $\Not^2 p$ & $\Not^3 p$ \\
\hline\hline
{\bf false} & {\bf true} & {\bf true} & {\bf false}\\
\hline
\end{tabular}
\end{center}
\vspace{- 1em}
\caption{A N$^4$ interpretation falsifying 
         $\Not p \Or p$}\label{fig:excluded-middle-falsifying-interpretation} 
\end{figure}  

Although $F \Or \Not F$, $\Not F \Or \Not^2 F$, and $\Not^2 F \Or \Not^3 F$
are true in all N$^4$ interpretations 
(Proposition \ref{prop:excluded-middle}), 
$F \Or \Not^3 F$ might be false in some N$^4$ interpretations. This is
for example the case of $F = p$ in the N$^4$ interpretation of Figure
\ref{fig:excluded-middle-falsifying-interpretation}. 

\begin{proposition}\label{prop:p-implies-not-not-p}~
If $F$ is in prefix negation form, for all $k \in \Nat$
$F \neq \Not^{2k+1} G$, and $val_{\cal I, V}(F) = {\bf true}$, 
then $val_{\cal I, V}(\Not^2 F) = {\bf true}$.
\end{proposition}

In N$^4$, implications are defined in terms of negation and
disjunction. In contrast to classical logic, in N$^4$ not all
disjunctions are expressible in terms of implications.   
For some formulas $F_1$ and $F_2$, {\rm N}$^4$
interpretations $\cal I$, and variable assignments $\cal V$, 
$val_{\cal I, V}((\Not F_1 \Implies F_2)) =$ {\bf true}
and 
$val_{\cal I, V}((F_1 \Or F_2)) =$ {\bf false}. This is the case,
e.g.\ if $F_1$ and $F_2$ are propositional variables and if $\cal I$
evaluates $F_1$ and $F_2$ as shown on Figure \ref{fig:proof}. 

\begin{figure}
\begin{center}
\begin{tabular}{|c|c|c|c|}
\hline
$F_1$       & $\Not^2 F_1$ & $F_2$       & $\Not^2 F_2$ \\
\hline\hline
{\bf false} & {\bf true}   & {\bf false} & {\bf true}    \\
\hline
{\bf false} & {\bf true}   & {\bf false} & {\bf false}    \\
\hline\end{tabular}
\end{center}
\vspace{- 1em}
\caption{A N$^4$ interpretation s.t.\ $(\Not F_1 \Implies F_2)$ is
         true and $(F_1 \Or F_2)$ false}\label{fig:proof} 
\end{figure}  

\begin{example}
Let $F_1 = (\Not p \Implies p)$. According to Definition
\ref{def:formula-valuation}, in N$^4$ $F_1$ is logically equivalent to 
$(\Not^2 p \Or p)$. Figure \ref{fg:models-1} gives
the possible valuations of $p$, $\Not p$, $\Not^2 p$ and $\Not^3 p$ in N$^4$
interpretations in which $F_1$ and $F_2$ are true.
In classical logic, only the first of these valuations is
possible.
\end{example}

\begin{figure}
\begin{center}
\begin{tabular}{|c|c|c|c|}
\hline
$p$         & $\Not p$    & $\Not^2 p$ & $\Not^3 p$ \\
\hline\hline
{\bf true} & {\bf false} & {\bf true}  & {\bf false}\\
\hline
{\bf false} & {\bf true} & {\bf true} & {\bf false}\\
\hline
\end{tabular}
\end{center}
\vspace{- 1em}
\caption{N$^4$ interpretations satisfying $(\Not p \Implies p)$}\label{fg:models-1}
\end{figure} 

\begin{example}
Let 
${\cal S}_2 = \{(\Not a \Implies b), (\Not b \Implies a)\}$. 
In N$^4$, 
${\cal S}_2$ is logically equivalent to 
$\{(\Not^2 a \Or b), (\Not^2 b \Or a)\}$. 
Figure \ref{fg:models-2} gives
the two possible valuations of $a$, $\Not^2 a$, 
$b$, and $\Not^2 b$ in N$^4$ interpretations in which 
${\cal S}_2$ is true. 
\end{example}

\begin{figure}
\begin{center}
\begin{tabular}{|c|c|c|c|c|c|c|c|}
\hline
$a$         & $\Not^2 a$  & $b$         & $\Not^2 b$  \\
\hline\hline
{\bf true}  & {\bf true}  & {\bf true}  & {\bf true}  \\
\hline
{\bf true}  & {\bf true}  & {\bf false} & {\bf true}  \\
\hline
{\bf true}  & {\bf true}  & {\bf false} & {\bf false} \\
\hline
{\bf false} & {\bf true}  & {\bf true}  & {\bf true}  \\
\hline
{\bf false} & {\bf true}  & {\bf false} & {\bf true}  \\
\hline
{\bf false} & {\bf false} & {\bf true}  & {\bf true}  \\
\hline
\end{tabular}
\end{center}
\vspace{- 1em}
\caption{N$^4$ interpretations satisfying 
         ${\cal S}_2 = \{(\Not a \Implies b), (\Not b \Implies a)\}$}\label{fg:models-2} 
\end{figure} 

\begin{example}
Let ${\cal S}_3 = \{(\Not p \Implies p), (p \Implies p)\}$. 
In N$^4$, ${\cal S}_3$ is logically equivalent to 
$\{(\Not^2 p \Or p), (\Not p \Or p)\}$. 
Figure \ref{fg:models-3} gives
the possible valuations of $p$, $\Not p$, and $\Not^2 p$  
in N$^4$ interpretations in which ${\cal S}_3$ is true. 
\end{example} 

\begin{figure}
\begin{center}
\begin{tabular}{|c|c|c|}
\hline
$p$         & $\Not p$    & $\Not^2 p$ \\
\hline\hline
{\bf true}  & {\bf false} & {\bf true} \\
\hline
{\bf false} & {\bf true}  & {\bf true} \\
\hline

\end{tabular}
\end{center}
\vspace{- 1em}
\caption{N$^4$ interpretation satisfying 
         ${\cal S}_3 = \{(\Not p \Implies p), (p \Implies p)\}$}\label{fg:models-3}
\end{figure} 

While a classical logic interpretation can be seen as N$^4$
interpretations, some N$^4$ interpretations have
no counterparts in classical logic. Such N$^4$ interpretations are
conveniently characterized as follows. 

\begin{definition}[In/Complete N$^4$ Interpretation]\label{def:complete-interpretation}
A {\rm N}$^4$ interpretation $\cal I$ of $\cal L$ is {\em
$F$-incomplete}, if for some variable assignment $\cal V$ with
respect to $\cal I$ $val_{\cal I, V}(F) \neq val_{\cal I, V}(\Not^2
F)$. Otherwise, it is {\em $F$-complete}.
A {\rm N}$^4$ interpretation is {\em incomplete}, if
it is $F$-incomplete for some formula $F$. Otherwise, it is {\em complete}. 
\end{definition}

\begin{proposition}\label{prop:complete-equivalents}
The following assertions are equivalent: 
\begin{enumerate}
   \item $\cal I$ induces a classical logic interpretation.
   \item $\cal I$ is complete.
   \item For all atoms $A$, $\cal I$ is $A$-complete. 
\end{enumerate}
\end{proposition}

\begin{definition}[N$^4$ Model]\label{def-model}
$\cal I$ is a {\em N$^4$ model} of $F$, if $val_{\cal I, V}(F) =$ {\bf
true} for some variable assignment $\cal V$ with respect to $\cal I$. 
A formula is {\em N$^4$ satisfiable} if it has a {\rm N}$^4$ model.
A formula is {\em N$^4$ falsifiable,} if there exists a {\rm N}$^4$
  interpretation in which this formula is false. 
\end{definition}

``$\cal I$ is a N$^4$ model of $F$'' will be noted ${\cal I}
\models_{{\rm N}^4} F$. ``$F_2$ logically
follows from $F_1$ in N$^4$'' will be noted $F_1 \models_{{\rm N}^4} F_2$.

\section{N$^4$ Herbrand Interpretations}

The classical definitions of the Herbrand base 
and Herbrand interpretations generated by a subset of the Herbrand base
extend straightforwardly to N$^4$.

\begin{definition}[N$^4$ Herbrand Interpretation]\label{herbrand-interpretation}
Let $U_{\cal L}$ denote the Herbrand universe of $\cal L$, i.e.\ the
set of all ground terms of $\cal L$. 
A {\rm N}$^4$ interpretation ${\cal I} = (D, val)$ is a {\rm N}$^4$ Herbrand
interpretation if
\begin{enumerate}
   \item $D = U_{\cal L}$
   \item For all $c \in {\it Const}_{\cal L}$, $val(c) = c$.
   \item For all $n \in \Nat \setminus \{0\}$, 
         $f \in {\it Fun}^0_{\cal L}$, 
         and $(t_1, \ldots, t_n) \in U_{\cal L}^{~n}, 
         val(f)(t_1, \ldots, t_n) = f(t_1, \ldots, t_n)$.
\end{enumerate}
\end{definition}

\begin{definition}[N$^4$ Herbrand Base]
The {\em N$^4$ Herbrand base} $B^2_{\cal L}$  of $\cal L$ is the set of
all positive ground {\rm N}$^4$ literals of $\cal L$. $M \subseteq
B^2_{\cal L}$ is {\em closed} if for all atom $A \in M$, 
$\Not^2 A \in M$.
\end{definition}

Thus, if $B_{\cal L}$ denotes the classical Herbrand base of a
first-order language ${\cal L}$, then in 
$B_{\cal L} \subseteq B^2_{\cal L}$ and (if ${\cal L}$ has some
predicate symbols) $B_{\cal L} \neq B^2_{\cal L}$. 

\begin{definition}[${\cal H}^2(M)$]\label{herbrand-interpretation-generated-by-a-set}
Let $M$ be a closed subset of $B^2_{\cal L}$. The unique {\rm N}$^4$ Herbrand
interpretation ${\cal H}^2(M)$ such that for all positive ground {\rm N}$^4$
literals $L \in B^2_{\cal L}$
\begin{center}
   ${\cal H}^2(M) \models_{{\rm N}^4} L$ iff $L \in M$ 
\end{center}
is the {\em N$^4$ Herbrand interpretation generated by} $M$. 
\end{definition}

The uniqueness of ${\cal H}^2(M)$ asserted in Definition 
\ref{herbrand-interpretation-generated-by-a-set} follows
immediately from Definition \ref{herbrand-interpretation}. 

The order on classical interpretations extends to N$^4$ interpretations.  

\begin{definition}[Order on N$^4$ Interpretations]\label{def:order}
Let ${\cal I}_1 = (D_1, val_1)$ and ${\cal I}_2 = (D_2, val_2)$ be two
{\rm N}$^4$ interpretations of $\cal L$.
${\cal I}_1$ is a {\em sub-interpretation} of ${\cal I}_2$, noted 
${\cal I}_1 \subseteq {\cal I}_2$, if
\begin{enumerate}
   \item $D_1 \subseteq D_2$,
   \item For all $c \in$ {\it Const}$_{\cal L}$ $val_1(c) = val_2(c)$,
   \item For all $n \in \Nat \setminus \{0\}$, $f \in$ 
         {\it Fun}$^n_{\cal L}$, and $(d_1, \ldots, d_n) \in D_1^{~n}$, \\
         $val_1(f)(d_1, \ldots, d_n) = val_2(f)(d_1, \ldots, d_n)$,
   \item For all $p \in$  {\it Rel}$^0_{\cal L}$, 
         $val_1(p) = val_2(p)$ and $val_1(\Not^2 p) = val_2(\Not^2 p)$. 
   \item For all $n \in \Nat \setminus \{0\}$ and $p \in$ 
         {\it Rel}$^n_{\cal L}$, 
         $val_1(p) = val_2(p) \cap D_1^{~n}$ and 
         $val_1(\Not^2 p) = val_2(\Not^2 p) \cap D_1^{~n}$.
\end{enumerate}
If in addition $D_1 \neq D_2$, then ${\cal I}_1$ is a {\em proper}
sub-interpretation of ${\cal I}_2$. 
\end{definition}

\begin{definition}[Intersection of N$^4$ Interpretations]\label{def:intersection}
Let $\{{\cal I}_k \mid k \in C\}$ be a collection of {\rm N}$^4$
interpretations of ${\cal L}$ such that
\begin{enumerate}
   \item ${\cal I}_k = (D_k, val_k)$.
   \item $C \neq \emptyset$. Let $k_0$ be an element of $C$.
   \item $D = \bigcap_{k \in C} D_k \neq \emptyset$.
   \item For all $c \in {\it Const}_{\cal L}$, $k \in C$, $l \in C$, 
         $val_k(c) = val_l(c)$.
   \item For all $n \in \Nat \setminus \{0\}$, 
         $f \in {\it Fun}^0_{\cal L}$, 
         $(d_1, \ldots, d_n) \in D$, $k \in C$, $l \in C$, 
         $val_k(f)(d_1, \ldots, d_n) = val_l(f)(d_1, \ldots, d_n)$.
\end{enumerate}
$\bigcap_{k \in C} {\cal I}_k = (D, val)$ is the {\rm N}$^4$ interpretation
with universe $D = \bigcap_{k \in C} D_k$ defined by:
\begin{enumerate}
   \item For all $c \in {\it Const}_{\cal L}$, $val(c) = val_{k_0}(c)$.
   \item For all $n \in \Nat \setminus \{0\}$, 
         $f \in {\it Fun}^0_{\cal L}$, 
         $(d_1, \ldots, d_n) \in D$, $k \in C$, $l \in C$, 
         $val(f)(d_1, \ldots, d_n) = val_{k_0}(f)(d_1, \ldots, d_n)$.
   \item For all $p \in {\it Rel}^0_{\cal L}$, 
         $val(p) =$ {\bf false} if $val_k(p)=$ {\bf false}
         for some $k \in C$, $val(p) =$ {\bf true} otherwise, and\\  
         $val(\Not^2 p) =$ {\bf false} if $val_k(\Not^2 p)$ {\bf false}
         for some $k \in C$, $val(\Not^2 p) =$ {\bf true}
         otherwise.
   \item For all $n \in \Nat$, $p \in {\it Rel}^n_{\cal L}$, 
         $val(p) = \bigcap_{k \in C} val_k(p)$ and 
         $val(\Not^2 p) = \bigcap_{k \in C} val_k(\Not^2 p)$.  
\end{enumerate}
\end{definition}

\begin{proposition}\label{prop:intersection}
Let $\{S_k \mid k \in C\}$ be a collection of subsets of the {\rm N}$^4$
Herbrand base $B^2_{\cal L}$. 
   \[{\cal H}^2(\bigcap_{k \in C} S_k) = 
     \bigcap_{k \in C} {\cal H}^2(S_k)\]

\end{proposition}

Let ${\cal P}_c(B^2_{\cal L})$ be the set of closed subsets of
$B^2_{\cal L}$. Since ${\cal P}_c(B^2_{\cal L})$ is obviously closed under
intersection and union, $({\cal P}_c(B^2_{\cal L}), \subseteq)$ is a
complete lattice. It follows from Propositions \ref{prop:intersection}
that $({\cal P}_c(B^2_{\cal L}), \subseteq)$ induces a complete lattice (the
order of which is also noted $\subseteq$) over the N$^4$ Herbrand
interpretations of ${\cal L}$. Referring to
this ordering, the minimal N$^4$ Herbrand models of a
set of formulas which is satisfiable over N$^4$ Herbrand interpretations
are well-defined. Thus, if $M$ is a closed subset of $B^2_{\cal L}$ and ${\cal S}$
is a set of formulas of ${\cal L}$, then ${\cal H}^2(M)$ is a minimal
N$^4$ Herbrand model of ${\cal S}$ iff:
\begin{enumerate}
   \item ${\cal H}^2(M) \models_{{\rm N}^4} {\cal S}$.
   \item For all $B$ closed subset of $M$ such that $B \neq M$, ${\cal H}^2(B)
         \not\models_{{\rm N}^4} {\cal S}$.
\end{enumerate}

The following characterization of minimal N$^4$ Herbrand models is used
in the next section.

\begin{proposition}\label{prop:minimal-model}
Let $\cal S$ be a set of formulas of $\cal L$ and $M$ a closed
subset of $B^2_{\cal L}$.
Let $\tilde{M} = \{\Not L \mid L \in B^2_{\cal L} \setminus M\}$.
${\cal H}^2_{\cal L}(M)$ is a minimal {\rm N}$^4$ Herbrand model of $\cal S$
iff
\begin{enumerate}
   \item ${\cal H}^2_{\cal L}(M) \models_{{\rm N}^4} {\cal S}$.
   \item For all $L \in M$, ${\cal S} \cup \tilde{M} \models_{{\rm N}^4} L$.
\end{enumerate}
\end{proposition}

\section{Minimal N$^4$ Herbrand Models of Normal Logic Programs}

Since double negations are not eliminated in N$^4$, the following 
interpretation of program clauses as formulas will be used. 

\begin{definition}[N$^4$ clausal form]\label{def:clausal-form} The {\em N$^4$ clause
associated with a general program clause} $A \If B_1, \ldots, B_n$
is the (closed) formula 
\begin{center}
$\forall x_1 \ldots \forall x_k 
~( \ldots (A \Or \Not B_1) \Or \ldots \Or \Not B_n)$ 
\end{center}
where $x_1, \ldots, x_k$ are the variables occurring in the literals $A, B_1,
\ldots,$ and $B_n$.
A {\rm N}$^4$ interpretation $\cal I$ {\em satisfies} a general
program clause $C$, if $\cal I$ is a {\rm N}$^4$  model of the 
{\rm N}$^4$ clause associated with $C$. 
Otherwise, it {\em falsifies} it. 
A interpretation $\cal I$ {\em satisfies} (or is a {\em {\rm N}$^4$ model}) of
a normal logic program, if it satisfies all its program clauses. 
Otherwise, it {\em falsifies} it. 
\end{definition}
Thus, the N$^4$ clause associated with the program clause 
   $a(x, y) \If b(x, z),$ $\Not c(z), d(y)$
is the formula
   $(\star)$~ $\forall x \forall y \forall z~ 
(((a(x, y) \Or \Not b(x, z)) \Or \Not^2 c(z)) \Or \Not d(y))$. 
Note that, in $(\star)$, double negations are not eliminated. Note also that 
$(\star)$ is logically equivalent (in N$^4$ and in classical logic) to 
   $\forall x \forall y \forall z~ 
(((b(x, z) \And \Not c(z)) \And d(y)) \Implies a(x, y))$.

Every normal logic program has a N$^4$ Herbrand model, since 
${\cal H}(B^2_{\cal L})$ is a model of every normal logic
program. Indeed, ${\cal H}(B^2_{\cal L})$ satisfies every N$^4$ clause
associated with 
a general program clause, because such a clause contains at
least one positive N$^4$ literal. Note that the classical minimal
Herbrand model of a positive logic program corresponds to 
its (unique) minimal N$^4$ Herbrand model. 

The following examples suggest that complete minimal N$^4$ Herbrand models 
might convey a logic program's intuitive meaning. The first two
examples are odd, resp.\ even length 
recursion cycles through negation. 

\begin{example}\label{ex-p-not-p}
Let ${\cal P}_1 = \{p \If \Not p\}$. In N$^4$, ${\cal P}_1$ is
logically equivalent to  ${\cal S}_1 =  \{(\Not^2 p \Or p)\}$. 
The unique minimal N$^4$ Herbrand model of ${\cal P}_1$ is
${\cal H}^2_{\cal L}(\{\Not^2 p\})$.
Figure \ref{fg:models-4} gives the valuations of $p$, $\Not
p$, and $\Not^2 p$ in this model 
(compare with Figure \ref{fg:models-1}). 
Note that ${\cal H}^2_{\cal L}(\{\Not^2 p\})$ is incomplete. 
\end{example}

\begin{figure}
\begin{center}
\begin{tabular}{|c|c|c|}
\hline
$p$         & $\Not p$    & $\Not^2 p$ \\
\hline\hline
{\bf false} & {\bf false} & {\bf true} \\
\hline
\end{tabular}
\end{center}
\vspace{- 1em}
\caption{Minimal N$^4$ model of $p \If \Not p$}\label{fg:models-4} 
\end{figure}  

\begin{example}
Let ${\cal P}_2 = \{a \If \Not b  ~;~ b \If \Not a\}$. 
In N$^4$ , ${\cal P}_2$ is logically equivalent to 
${\cal S}_2 =  \{ (\Not^2 b \Or a), (\Not^2 a \Or b)\}$. 
The minimal N$^4$ Herbrand models of ${\cal P}_2$ are 
${\cal H}^2_{\cal L}(\{a, \Not^2 a\})$, 
${\cal H}^2_{\cal L}(\{\Not^2 a, \Not^2 b\})$, 
and
${\cal H}^2_{\cal L}(\{b, \Not^2 b\})$. 
Figure \ref{fg:models-5} gives the valuations of 
the N$^4$ literals in these models
(compare with Figure \ref{fg:models-2}).
${\cal H}^2_{\cal L}(\{\Not^2 a, \Not^2 b\})$ is
incomplete.
\end{example}

\begin{figure}
\begin{center}
\begin{tabular}{|c|c|c|c|c|c|}
\hline
$a$         & $\Not a$    & $\Not^2 a$ & $b$         & $\Not b$    & $\Not^2 b$ \\
\hline\hline
{\bf true}  & {\bf false} & {\bf true}    & {\bf false} & {\bf true}  & {\bf false}\\
\hline
{\bf false} & {\bf false} & {\bf true}    & {\bf false} & {\bf false} & {\bf true} \\
\hline
{\bf false} & {\bf true} & {\bf false}    & {\bf true}  & {\bf false} & {\bf true} \\
\hline
\end{tabular}
\end{center}
\vspace{- 1em}
\caption{Minimal N$^4$ models of ${\cal P} = \{b \If \Not a ~;~ a \If
  \Not b\}$}\label{fg:models-5} 
\end{figure} 

\begin{example}
Let ${\cal P}_3 = {\cal P}_1 \cup {\cal P}_2 = 
\{p \If \Not p, \Not a ~;~ a \If \Not b  ~;~ b \If \Not a\}$. 
The minimal N$^4$ Herbrand models of ${\cal P}_3$ are 
${\cal H}^2_{\cal L}(\{a, \Not^2 a\})$, 
${\cal H}^2_{\cal L}(\{\Not^2 a, \Not^2 b\})$, 
and
${\cal H}^2_{\cal L}(\{\Not^2 p, b, \Not^2 b\})$. Compare with the previous
examples. 
\end{example}

\begin{example}
Let ${\cal P}_4 = {\cal P}_1 \cup \{p \If p\}$. In N$^4$, ${\cal P}_4$
is logically equivalent to ${\cal S}_4 = \{(\Not^2 p \Or p), 
(\Not p \Or p)\}$. It follows from Proposition \ref{prop:excluded-middle} (2) that 
${\cal S}_4$ is logically equivalent to ${\cal S}_1 = \{(\Not^2 p \Or
p)\}$. Thus, ${\cal P}_1$ and ${\cal P}_4$ have the same
minimal N$^4$ Herbrand model ${\cal H}^2_{\cal L}(\{\Not^2 p\})$. 
\end{example}

\begin{proposition}\label{prop:simpl}
Let $\cal S$ be a (possibly infinite) set of ground (general) program
clauses. If $M$ is a closed subset of $B^2_{\cal L}$, 
let ${\rm Simp}_M({\cal S})$ denote
the set of ground general program clauses obtained from $\cal S$ as
follows: 
\begin{enumerate}
   \item First delete all clauses whose bodies contain some negative
         literal $\Not A$ with $A \in M$.
   \item Second, delete the negative literals from the bodies
         of the remaining clauses. 
\end{enumerate}
Let $A$ be a ground atom. 
${\cal S} \cup \tilde{M} \models_{{\rm N}^4} A$ 
iff 
${\rm Simp}_M({\cal S}) \models A$.
\end{proposition}

Proposition \ref{prop:simpl} does not hold in classical
logic. Consider for example ${\cal P}_1 = \{p \If \Not p\}$. Assume
that $p$ is the only predicate symbol of $\cal L$ and let $M =
\{p, \Not^2 p\}$. In classical logic  
${\cal P}_1 \cup \tilde{M} = {\cal P}_1 \models p$ but 
${\rm Simp}_{M}({\cal P}_1) = \emptyset \not\models p$. 

\begin{proposition}\label{prop:stable-iff-complete-minimal} 
Let $\cal P$ be a normal logic program. 
A {\rm N}$^4$ Herbrand model of $\cal P$ is stable iff it is
complete and minimal. 
\end{proposition}

\section{Perspectives and Related Work}

The approach presented here seems to enjoy many of the strong and weak
principles of \cite{dix93a,dix93b}. E.g.\ ``Cut'', ``Cautious
Monotonicity'', and the ``Principle of Partial
Evaluation'' result directly from the the classical-style evaluation
function (Definition \ref{def:formula-valuation}), ``Relevance'' from 
N$^4$ treatment of double negations and from 
model minimality (Proposition \ref{prop:simpl}). 
This deserves deeper investigations.  

The model theory of N$^4$ presented in this paper needs to be
complemented with a proof theory. First investigations 
indicate that natural deduction and the tableau method well 
adapt to N$^4$. A tableau method
for N$^4$ would provide with a basis for defining a fixpoint-like
generation of the minimal N$^4$ Herbrand models of a normal logic
program. Also, it would be useful for program development to have at
disposal a backward reasoning method able to detect whether, for some
instance $G\sigma$ of a goal $G$, a logic program has a
$G\sigma$-incomplete minimal N$^4$ Herbrand model. 

Publications on the semantics of normal logic programs are numerous --
cf. the surveys 
\cite{bidoit91,dix93a,dix93b,dix93c,apt94,dix98}. For space reasons,
these publications cannot be discussed here in detail. Most of them
can be roughly classified in ad hoc definitions of models (such as
\cite{gelfond88}) for (restricted or unrestricted) normal logic
programs, forward reasoning methods for computing models, 
approaches referring to non-standard logics (often three-valued
logics), and approaches based on program transformations. The approach
presented here is of the first and third types. 
Its particularities are that it is based upon a notion of minimal
Herbrand models and that it refers to a nonstandard logic rather close
to classical logic. Note interesting similarites with the
transformation-based approach of \cite{drabent91,wallace93}.
Note also that N$^4$ can be seen as a four-valued logic (the truth
values of which can be read ``true'', ``false'', ``required'', and ``not
required'').  

Aspects of the work presented here have been inspired from
\cite{inoue96,niemelae96} as follows. The interpretation of program clauses as 
N$^4$ formulas (Definition \ref{def:clausal-form}) is reminiscent of 
their processing in \cite{inoue96}. The characterization of minimal
N$^4$ Herbrand models (Proposition \ref{prop:minimal-model}) is
an adaptation to N$^4$ of a result given in \cite{niemelae96} for
classical logic. 

\bigskip\section*{Appendix: Proofs}

\bigskip\noindent\textit{Proof of Proposition \ref{prop:replacement}:}

\begin{enumerate}    
    \item[1.] $val_{\cal I, V}(\Falsum) =$ {\bf false} iff 
               (Def.\ \ref{def:formula-valuation} (5))
              $val_{\cal I, V}(\Falsum) \neq$ {\bf true}. 
              This holds, since none of the cases of Def.\
              \ref{def:formula-valuation} give rise to derive 
              $val_{\cal I, V}(\Falsum) =$ {\bf true}.\\
              $val_{\cal I, V}(\Verum) = 
               val_{\cal I, V}(\Not \Falsum) =$ {\bf true} by 
              Def.\ \ref{def:formula-valuation} (2.1).
   \item[2-10.] Each statement follows from the
                corresponding property of the meta-language 
                in which Definition \ref{def:formula-valuation} 
                is expressed.
                \hfill \rule{1.5mm}{1.5mm} 
\end{enumerate}

\bigskip\noindent\textit{Proof of Proposition \ref{prop:p-not-p}:}

\bigskip\noindent
The proof is by induction on the structure of $F$. Let $k \in \Nat$
and ${\rm IH}(G)$ denote: 
``$val_{\cal I, V}(G) \neq val_{\cal I, V}(\Not G)$''.

\bigskip\noindent
{\bf Basis cases:}

\bigskip\noindent
$F$ is an atom or $F = \Falsum$. ${\rm IH}(\Falsum)$ holds since Def.\
\ref{def:formula-valuation} (2.1) and Prop.\
\ref{prop:replacement} (1).

\bigskip\noindent
{\bf Induction cases:}
\begin{enumerate}
   \item $F = (F_1 \And F_2)$. Assume ${\rm IH}(F_1)$ and 
         ${\rm IH}(F_2)$ (ind.\  hyp.). ${\rm IH}(F)$ follows from Def.\
         \ref{def:formula-valuation} (1.2, 2.2), ${\rm IH}(F_1)$ and 
         ${\rm IH}(F_2)$. 

   \item $F = (F_1 \Or F_2)$. Assume ${\rm IH}(F_1)$ and 
         ${\rm IH}(F_2)$ (ind.\  hyp.).${\rm IH}(F)$ follows from Def.\
         \ref{def:formula-valuation} (1.3, 2.3), ${\rm IH}(F_1)$ and 
         ${\rm IH}(F_2)$.

   \item $F = \forall x F_1$. Assume ${\rm IH}(F_1)$ 
          (ind.\  hyp.). ${\rm IH}(F)$ follows from Def.\
         \ref{def:formula-valuation} (1.4, 2.4) and ${\rm IH}(F_1)$. 

   \item $F = \exists x F_1$. Assume ${\rm IH}(F_1)$ 
          (ind.\  hyp.). ${\rm IH}(F)$ follows from Def.\
         \ref{def:formula-valuation} (1.5, 2.5) and ${\rm IH}(F_1)$.
         \hfill \rule{1.5mm}{1.5mm}
\end{enumerate}

\bigskip\noindent\textit{Proof of Proposition \ref{prop:excluded-middle}:}

\begin{enumerate}
   \item $val_{\cal I, V}(\Not^4 F) = {\bf true}$ iff 
             (Def. \ref{def:formula-valuation} (4))
         $val_{\cal I, V}(\Not^3 F) \neq {\bf true}$ iff
             (Def. \ref{def:formula-valuation} (4)) 
         $val_{\cal I, V}(\Not^2 F) = {\bf true}$. 
 
   \item The proof is by induction on the structure of $F$.
         Let ${\rm IH}(G)$ denote:
             ``$val_{\cal I, V}((G \Or \Not G)) =
               val_{\cal I, V}((\Not G \Or \Not^2 G)) =
               val_{\cal I, V}((\Not^2 G \Or \Not^3 G)) =
               {\bf true}$.''

         {\bf Basis cases:}

                  \noindent
                  $F$ is an atom or $F = \Falsum$. \\
                  $val_{\cal I, V}((F \Or \Not F)) = {\bf true}$ by
                     Def.\ \ref{def:formula-valuation} (1.3, 2.1).\\
                  $val_{\cal I, V}((\Not F \Or \Not^2 F)) = {\bf true}$ by
                     Def.\ \ref{def:formula-valuation} (1.3, 3.1). \\
                  $val_{\cal I, V}((\Not^2 F \Or \Not^3 F)) = {\bf true}$ by
                     Def.\ \ref{def:formula-valuation} (1.3, 4). 

         {\bf Induction cases:} 
         \begin{enumerate}

            \item[1.] $F = \Not F_1$. Assume ${\rm IH}(F_1)$ (ind.\  hyp.). 

                      $val_{\cal I, V}((F \Or \Not F)) = 
                       val_{\cal I, V}((\Not F_1 \Or \Not^2 F_1)) =
                       {\bf true}$
                          (${\rm IH}(F_1)$).

                      $val_{\cal I, V}((\Not F \Or \Not^2 F)) = 
                       val_{\cal I, V}((\Not^2 F_1 \Or \Not^3 F_1)) = 
                       {\bf true}$
                          (${\rm IH}(F_1)$).

                      By Def. \ref{def:formula-valuation} (3.3), 
                      Prop.\ \ref{prop:replacement} (1), and 
                      ${\rm IH}(F_1)$:\\
                      $val_{\cal I, V}((\Not^2 F \Or \Not^3 F)) =
                       val_{\cal I, V}((\Not^3 F_1 \Or \Not^4 F_1)) =\\
                       val_{\cal I, V}(\Not^2 (\Not F_1 \Or \Not^2 F_1)) =
                       val_{\cal I, V}(\Not^2 \Verum) =$ {\bf true}. 

            \item[2.] $F = (F_1 \And F_2)$. Assume ${\rm IH}(F_1)$ and
                      ${\rm IH}(F_2)$ (ind.\  hyp.).

                      By Def. \ref{def:formula-valuation} (3.3), 
                      Prop.\ \ref{prop:replacement},   
                      ${\rm IH}(F_1)$, and ${\rm IH}(F_2)$

                      $val_{\cal I, V}((F \Or \Not F)) = 
                       val_{\cal I, V}(((F_1 \And F_2) \Or \Not (F_1 \And F_2))) =\\ 
                       val_{\cal I, V}( ( ((F_1 \Or \Not F_1) \Or F_2)
                                         \And
                                         ((F_2 \Or \Not F_2) \Or F_1))) =\\ 
                       val_{\cal I, V}(((\Verum \Or F_2) 
                                        \And 
                                        (\Verum \Or F_1)))) =$ {\bf true}.
                      
                       $val_{\cal I, V}((\Not F \Or \Not^2 F)) = 
                       val_{\cal I, V}((\Not (F_1 \And F_2) 
                                        \Or 
                                        \Not^2 (F_1 \And F_2))) =\\
                        val_{\cal I, V}(((\Not F_1 \Or \Not^2 F_1) 
                                         \And
                                         (\Not F_2 \Or \Not^2 F_2))) =
                        val_{\cal I, V}((\Verum \And \Verum)) =$ 
                      {\bf true}.

                      $val_{\cal I, V}((\Not^2 F \Or \Not^3 F)) = 
                        val_{\cal I, V}((\Not^2 (F_1 \And F_2) 
                                        \Or 
                                        \Not^3 (F_1 \And F_2))) =\\
                        val_{\cal I, V}(((\Not^2 F_1 \Or \Not^3 F_1) 
                                         \And
                                         (\Not^2 F_2 \Or \Not^3 F_2))) =
                        val_{\cal I, V}((\Verum \And \Verum)) =$ 
                      {\bf true}.

            \item[3.] $F = (F_1 \Or F_2)$. 
                      The proof is similar to those of the preceding case.

            \item[6.] $F = \forall x F_1$. Assume ${\rm IH}(F_1)$ (ind. hyp.). 
                      By Def. \ref{def:formula-valuation} (3.3), 
                      Prop.\ \ref{prop:replacement}, and  
                      ${\rm IH}(F_1)$: 

                      $val_{\cal I, V}((F \Or \Not F)) = 
                       val_{\cal I, V}((\forall x F_1 \Or \Not \forall x F_1)) =
                       val_{\cal I, V}(\forall x (F_1 \Not F_1)) =
                       val_{\cal I, V}(\forall x \Verum) =
                       val_{\cal I, V}(\Verum) =$
                      {\bf true}. 

                     $val_{\cal I, V}((\Not F \Or \Not^2 F)) = 
                       val_{\cal I, V}((\Not \forall x F_1 \Or \Not^2 \forall x F_1)) =\\
                       val_{\cal I, V}(\forall x (\Not F_1 \Or \Not^2 F_1)) =
                       val_{\cal I, V}(\forall x \Verum) =
                       val_{\cal I, V}(\Verum) =$
                      {\bf true}. 

                     $val_{\cal I, V}((\Not^2 F \Or \Not^3 F)) = 
                       val_{\cal I, V}((\Not \forall x F_1 \Or \Not^3 \forall x F_1)) =\\
                       val_{\cal I, V}(\forall x (\Not^2 F_1 \Or \Not^3 F_1)) =
                       val_{\cal I, V}(\forall x \Verum) =
                       val_{\cal I, V}(\Verum) =$
                      {\bf true}. 

            \item[7.] $F = \exists x F_1$. 
                      The proof is similar to those of the preceding
                      case.
                      \hfill \rule{1.5mm}{1.5mm}
         \end{enumerate} 
\end{enumerate}    

\bigskip\noindent\textit{Proof of Proposition \ref{prop:p-implies-not-not-p}:}

\bigskip\noindent
First note that $val_{\cal I, V}(\Not^2 \Verum)$ = {\bf true} since Def.\
\ref{def:formula-valuation} (4, 3.1). 
Assume $val _{\cal I, V}(F) =$ {\bf true}. Then, by Prop.\
\ref{prop:replacement} (8) 
$val _{\cal I, V}(\Not^2 F) = 
val _{\cal I, V}(\Not^2 \Verum)$. By Def.\ \ref{def:formula-valuation}
(4, 2.1), $val _{\cal I, V}(\Not^2 \Verum) =$ {\bf true}. 
Hence, $val _{\cal I, V}(\Not^2 F) = $ {\bf true}.  
\hfill \rule{1.5mm}{1.5mm}

\bigskip\noindent\textit{Proof of Proposition \ref{prop:complete-equivalents}:}

\bigskip\noindent
$1 \Equiv 2$: $2$ is a rephrasing of $1$.

\noindent
$2 \Implies 3$: By definition of complete and
$A$-complete N$^4$ interpretations.

\noindent
$3 \Implies 2$: Let $\cal I$ be a N$^4$ interpretation which is $A$-complete
for all atoms $A$. 
The proof is by induction on the structure of $F$.

\bigskip\noindent
{\bf Basis cases:}
\begin{enumerate}
   \item $F = \Falsum$.  
         By Prop.\ \ref{prop:replacement} (1), 
         $val_{\cal I, V}(\Falsum) =$ {\bf false}
         and 
         $val_{\cal I, V}(\Not \Falsum) = val_{\cal I, V}(\Verum) =$ {\bf true}.
         Therefore, $\cal I$ is $\Falsum$-complete. 

   \item $F$ is an atom. $\cal I$ is $F$-complete, since by hypothesis, 
         it is $A$-complete for all atoms $A$.
\end{enumerate}
\noindent
{\bf Induction cases:}
\begin{enumerate}
   \item $F = (F_1 \And F_2)$. Assume that $\cal I$ is $F_1$-complete
     and $F_2$-complete (ind.\ hyp.). By Def.\ \ref{def:formula-valuation}
     (1.2, 3.2), $val_{\cal I, V}(F) = val_{\cal I, V}(\Not^2 F)$.

   \item $F = (F_1 \Or F_2)$. Assume that $\cal I$ is $F_1$-complete
         and $F_2$-complete (ind.\ hyp.). By Def.\ \ref{def:formula-valuation}
         (1.3, 3.3), $val_{\cal I, V}(F) = val_{\cal I, V}(\Not^2 F)$.

   \item $F = \forall x F_1$. Assume that $\cal I$ is $F_1$-complete
         (ind.\ hyp.). By Prop.\ \ref{prop:replacement}
         (9, 10), $val_{\cal I, V}(F) = val_{\cal I, V}(\Not^2 F)$.

   \item $F = \exists x F_1$. Assume that $\cal I$ is $F_1$-complete
         (ind.\ hyp.). By Prop.\ \ref{prop:replacement}
         (9, 10), $val_{\cal I, V}(F) = val_{\cal I, V}(\Not^2 F)$.

   \item $F = \Not F_1$. Assume that $\cal I$ is $F_1$-complete
         (ind.\ hyp.). By Prop.\ \ref{prop:replacement} (8) 
         $val_{\cal I, V}(F) = 
          val_{\cal I, V}(\Not F_1) = 
          val_{\cal I, V}(\Not^3 F_1) = 
          val_{\cal I, V}(\Not^2 F)$. 
          \hfill \rule{1.5mm}{1.5mm}
\end{enumerate}

\bigskip\noindent\textit{Proof of Proposition \ref{prop:intersection}:}

\bigskip\noindent
As its classical logic counterparts, the result follows directly from
the definition of the intersection of interpretations (Def.\ \ref{def:intersection}).
\hfill \rule{1.5mm}{1.5mm}

\bigskip\noindent\textit{Proof of Proposition \ref{prop:minimal-model}:}

\bigskip\noindent
{\bf Necessary condition:} 
Assume that ${\cal H}^2(M)$ is a minimal {\rm N}$^4$ Herbrand model of
$\cal S$. Thus, 1 holds. If $M = \emptyset$, then 2 holds
trivially. Otherwise, let $L \in M$. Let $\tilde{L} = \Not L$. If 
${\cal S} \cup \tilde{M} \not\models_{{\rm N}^4} L$, then  
${\cal S} \cup \tilde{M} \cup \{\tilde{L}\}$ has a {\rm N}$^4$
Herbrand model, hence also a minimal {\rm N}$^4$ Herbrand model, say 
${\cal H}^2(N)$. By definition of ${\cal H}^2(N)$, 
${\cal H}^2(N) \models_{{\rm N}^4} \tilde{L}$. Therefore, 
${\cal H}^2(N) \not\models_{{\rm N}^4} L$
(Prop.\ \ref{prop:p-not-p}), i.e.\ $L \not\in N$. Since 
${\cal H}^2(N) \models_{{\rm N}^4} {\cal S} \cup \tilde{M}$, $N \subseteq
M$. Since $L \in M \setminus N$, $N \neq M$. This contradict the
minimality of ${\cal H}^2(M)$ since 
${\cal H}^2(N) \models_{{\rm N}^4} {\cal S}$.
Therefore, for all $L \in M$, ${\cal S} \cup \tilde{M}
\models_{{\rm N}^4} L$, i.e.\ 2 holds.

\bigskip\noindent
{\bf Sufficient condition:}
Assume that 1 and 2 hold. If ${\cal H}^2(M)$ is not a minimal N$^4$ Herbrand
model of $\cal S$, then there exists a closed subset $N$ of $M$ such
that $N \neq M$ and ${\cal H}^2(N)$ is a minimal N$^4$ Herbrand model of $\cal S$. 
Let $\tilde{N} =  \{\Not L \mid L \in B^2_{\cal L} \setminus N\}$.
From the necessary condition, it follows that for all $L \in N$, 
${\cal S} \cup \tilde{N} \models_{{\rm N}^4} L$. Since $N \subset
M$, $N \neq M$, there exists $L \in M \setminus N$. By hypothesis 2, 
${\cal S} \cup \tilde{M} \models_{{\rm N}^4} L$. Since $N \subset
M$, $\tilde{M} \subset \tilde{N}$. Therefore, 
${\cal S} \cup \tilde{N} \models_{{\rm N}^4} L$. But by
definition, $L \in M \setminus N$. Therefore, 
$\tilde{L} \in \tilde{N}$. Thus, both $L$ and $\tilde{L}$ are
true in every model of ${\cal S} \cup \tilde{N}$, among others in 
${\cal H}^2(N)$. This contradicts Prop.\ \ref{prop:p-not-p}. Therefore, 
${\cal H}^2(M)$ is a minimal N$^4$ Herbrand model of $\cal S$.
\hfill \rule{1.5mm}{1.5mm}

\bigskip\noindent\textit{Proof of Proposition \ref{prop:simpl}:}

\bigskip\noindent
By definition of N$^4$ interpretations (Def.\ \ref{def:n3-interpretation}), 
$\Not^2 A \not\models_{{\rm N}^4} A$ for all ground atoms $A$. Therefore, 
${\cal S} \cup \tilde{M} \models_{{\rm N}^4} A$ for some ground atom $A$ 
iff 
for all N$^4$ interpretation $\cal I$ such that ${\cal I}
\not\models_{{\rm N}^4} M$, there exists a
program clause $C \in {\cal S}$ such that:
\begin{enumerate}
   \item $A$ is the head of $C$.
   \item For all positive body literal $B$ of $C$, 
         ${\cal I} \models_{{\rm N}^4} B$. 
   \item For all negative body literals $\Not B$ of $C$, 
         ${\cal I} \not\models_{{\rm N}^4} B$.
\end{enumerate}
Thus, 
${\cal S} \cup \tilde{M} \models_{{\rm N}^4} A$ for some ground atom $A$ 
iff 
${\rm Simp}_M({\cal S}) \models_{{\rm N}^4} A$. 
Since no negative literal occur in the program clauses in ${\rm Simp}_M({\cal S})$, 
${\rm Simp}_M({\cal S}) \models_{{\rm N}^4} A$ implies 
${\rm Simp}_M({\cal S}) \models A$. 
\hfill \rule{1.5mm}{1.5mm}

\bigskip\noindent\textit{Proof of Proposition \ref{prop:stable-iff-complete-minimal}:}

\bigskip\noindent
Let ${\rm Ground}({\cal P})$ denote the set of ground instances of
the program clauses of a normal logic program  $\cal P$. 
A stable model of $\cal P$ \cite{gelfond88} is a classical logic Herbrand
interpretation ${\cal H}_{\cal L}(M)$ ($M \subseteq B_{\cal L}$) such that, 
for all atoms $A$, 
$A \in M$ iff 
${\rm Simp}_M({\rm Ground}({\cal P})) \models A$.

\bigskip\noindent
If $M \subseteq B_{\cal L}$, let  
$\overline{M} = \{\Not A \mid A \in M\}$
and 
$\overline{\overline{M}} = \{\Not^2 A \mid A \in M\}$.

\noindent
{\bf Necessary condition:}
Let ${\cal H}_{\cal L}(M)$ be a stable model of $\cal P$ 
(i.e.\ $M \subseteq B_{\cal L}$). 
Since ${\cal H}_{\cal L}(M) \models \cal P$,
${\cal H}^2_{\cal L}(M \cup \overline{\overline{M}})$ is a complete
N$^4$ model of $\cal P$. 
Since ${\cal H}^2_{\cal L}(M \cup \overline{\overline{M}})$
is complete, 
$(\star)$
for all $A \in B_{\cal L}$, 
${\cal H}^2_{\cal L}(M \cup \overline{\overline{M}}) \models_{{\rm N}^4} \Not^2 A$ iff 
${\cal H}^2_{\cal L}(M \cup \overline{\overline{M}}) \models_{{\rm N}^4} A$. 
Let $A \in M$. 
Since ${\cal H}_{\cal L}(M)$ is a stable model of $\cal P$, 
${\rm Simp}_M({\rm Ground}({\cal P})) \models A$. 
Therefore, 
${\rm Simp}_M({\rm Ground}({\cal P})) \models_{{\rm N}^4} A$. 
It follows from $(\star)$ that 
for all $L \in M \cup
\overline{\overline{M}}$, 
${\rm Simp}_M({\rm Ground}({\cal P})) \models_{{\rm N}^4} L$. 
I.e.\ by Prop.\ \ref{prop:minimal-model}  
${\cal H}^2_{\cal L}(M)$ is a minimal N$^4$  model of $\cal P$.

\bigskip\noindent 
{\bf Sufficient condition:} 
Let ${\cal H}^2_{\cal L}(M)$ be a complete and minimal N$^4$ model of $\cal P$. Let
Let $N = M \cap B_{\cal L}$. Since ${\cal H}^2_{\cal L}(M)$ is complete, $M = N \cup
\overline{\overline{N}}$. Since ${\cal H}^2_{\cal L}(M)$ is a minimal
model of $\cal P$, by Prop.\ \ref{prop:minimal-model}, 
for all $A \in N$, 
${\cal P} \cup \tilde{M} \models_{{\rm N}^4} A$.
Therefore, 
for all $A \in N$, 
${\rm Simpl}_{M}({\rm Ground}({\cal P})) \models_{{\rm N}^4} A$.
By Prop.\ \ref{prop:simpl}, 
for all $A \in B_{\cal L}$, 
${\rm Simpl}_{M}({\rm Ground}({\cal P})) \cup \tilde{M} \models A$. 
I.e.\ ${\cal H}_{\cal L}(N)$ is a stable model of $\cal P$. 
\hfill \rule{1.5mm}{1.5mm}

\let\And=\goodoldAnd

\end{document}